# Thermodynamic stability criterion and fluctuation theory in nonextensive thermodynamics

Zheng Yahui [a,b,c], Du Jiulin [b], Liang Faku [c]

a: Department of Electronics and Communication Engineering, Henan Institute of Technology, Xinxiang 453003, China.
b: Department of Physics, School of Science, Tianjin University, Tianjin 300072, China.
c: Department of Physics, School of Science, Qiqihar University, Qiqihar 161006, China.

**Abstract**

We have constructed a nonextensive thermodynamic formalism consisting of two sets of parallel Legendre transformation structures in previous papers. One is the physical set and the other is the Lagrange set. In this paper we study the thermodynamic stability criterion with a dual interpretation of the thermodynamic relations and quantities. By recourse to the assumption that volume in nonextenstive system is nonadditive, we conclude that it is the physical pressure that is responsible for the mechanical balance between any two parts in a given nonextensive system. It is verified that in the physical set of transformation structures, the stability criterion can be expressed in terms of heat capacity and isothermal compressibility. We also discuss the fluctuation theory in nonextensive thermodynamics.

## 1. Introduction

It has been found that many complex systems, including the systems with long-range interactions, with fractal space-time and with non-Markov process, cannot be well described by the classical Boltzmann-Gibbs statistics. There are many examples which show that the classical statistics can be satisfactory only for describing the systems where the short-range interactions are dominant. In recent years, based on the q-entropy proposed by Tsallis in 1988 [1], nonextensive statistical mechanics (NSM) has been developed to describe the complex systems. By use of this new statistical theory, a type of power-law q-distribution functions can be studied. So far, NSM has been applied to many research fields in various complex systems, such as plasma physics [2-5], astrophysics [6–9], systems with anomalous diffusion [10,11], biological physics [12-14] and econophysics [15,16] etc.

Nonextensive statistics has made great progress in exploring and understanding the complex systems with power-law distributions, but further research is still needed in nonextensive thermodynamics so as to develop a complete formalism like those in traditional thermodynamics [17-23]. One of the long standing problems may be how to define temperature in nonextensive thermodynamics by the connection to the zeroth law of thermodynamics [24]. Generally, there have been two points about the definition of temperature in nonextensive thermodynamics. One point is how to define a homogeneous state-variable with temperature, the other point is how to apply it to a nonequilibrium complex system. Many authors have made some discussions about the first point in different situations. For example, Abe analyzed a Tsallis equilibrium in the statistical ensemble, defined a physical temperature and applied it to an ideal gas model [17,25,26].

Martinez et al [27] reconciled the nonextensive thermodynamics with the zeroth law by recourse to a modified Lagrange multiplier. Johal deduced an additive entropy expression by use of algebraic methods [28,29]. Chen and Ou [24] studied a normal Legendre relation based on the isothermal assumption and the probability independence assumption. Scarfone explored an addition rule of the energy level of microstate using algebraic analysis [30].

However, the second point has not aroused enough attention. In order to study the second point, we proposed the temperature duality assumption [31,32], by which two sets of parallel Legendre transformation structures were made, i.e., the 'physical' set (P-set) and 'Lagrange' set (L-set) [33]. The P-set formalism consists of the quantities manifesting an obvious nonextensive nature or non-measurability, while the L-set formalism is made with measurable thermodynamic quantities. In the each formalism, the thermodynamic relations and quantities are completely the same in the form as those in traditional thermodynamics. The two sets of thermodynamic formalisms can be linked to each other by the Tsallis factor, which apparently developed the idea of the temperature duality and made nonextensive thermodynamics easier to apply to a nonequilibrium complex system.

In this work, we will further discuss the thermodynamic stability criterion. For this purpose, this paper is constructed as follows. In section 2, we discuss the balance conditions of a nonextensive system with two different phases. In section 3, we discuss the stability criterion in the nonextensive thermodynamics. And in section 4, we study a generalization of the fluctuation theory in nonextensive thermodynamics. In section 5, we give the summaries and conclusions.

## 2. The equilibrium conditions of a nonextensive biphase system

The q-entropy of a nonextensive system is expressed [1] by

$$S_q = k \frac{\sum_i p_i^q - 1}{1-q} \equiv k \sum_i p_i \ln_q(\frac{1}{p_i}), \quad (1)$$

where $k$ is Boltzmann constant, $p_i$ is the probability of the $i$th microstate of the system, and $q$ is the nonextensive parameter whose difference from one describes the degree of nonextensivity. For convenience, we introduce the Tsallis factor, $c_q \equiv \sum_i p_i^q$.

We consider an isolated nonextensive system consisting of two different phases. In each phase, the system has its own internal energy. In nonextensive thermodynamics, if the internal energy may be assumed two different concepts [33], one may be physical internal energy (or so called P-energy), defined as

$$U_q^{(2)} \equiv \sum p_i^q \varepsilon_i, \quad (3)$$

while the other may be Lagrange internal energy (or so called L-energy), defined as

$$U_q^{(3)} \equiv \frac{\sum p_i^q \varepsilon_i'}{c_q}, \quad (4)$$

Notice that in the above two definitions the microstate energy levels are different from each other. Then in the two-phase nonextensive system marked by (1, 2) can satisfy different addition rules. They are, respectively [33],

$$\varepsilon_{(i,j)(1,2)} = \varepsilon_{i1} + \varepsilon_{j2} - (1-q)\beta' \varepsilon_{i1} \varepsilon_{j2}, \quad (5)$$

$$\varepsilon'_{(i,j)(1,2)} = \varepsilon'_{i1} + \varepsilon'_{j2} - (1-q)\frac{\beta}{c_q}[\varepsilon'_{i1} - U^{(3)}_{q1}][\varepsilon'_{j2} - U^{(3)}_{q2}], \tag{6}$$

where "1" and "2" represent different phases; "$i$" and "$j$" represent different microscopic states. $\beta$ is the Lagrange multiplier, while $\beta'$ is a generalized Lagrange multiplier, $\beta' = \beta/c_q$.

Corresponding to the equations (3) and (4), the first law of thermodynamics can be expressed as two different forms,

$$dU^{(2)}_q = T_q dS_q - P_q dV, \text{ and} \tag{7}$$

$$dU^{(3)}_q = T dS_q - P dV, \tag{8}$$

where the concept of temperature duality [33], i.e., the physical temperature $T_q$ and the Lagrange temperature $T$, gives that

$$T_q = \frac{1}{k\beta'}, \quad T = \frac{1}{k\beta}. \tag{9}$$

And they are connected to each other by the relation: $T_q = c_q T$. Correspondingly, the physical pressure $P_q$ and the Lagrange pressure $P$ are linked by the relation: $P_q = c_q P$.

Now that the nonextensive system consists of two different phases, the total q-entropy satisfies

$$S_{q1,2} = S_{q1} + S_{q2} + \frac{1-q}{k} S_{q1} S_{q2}. \tag{10}$$

In the two internal energy descriptions (3) and (4), the P-energy addition rule is given by

$$U^{(2)}_{q1,2} = c_{q2} U^{(2)}_{q1} + c_{q1} U^{(2)}_{q2} - (1-q)\beta' U^{(2)}_{q1} U^{(2)}_{q2}, \tag{11}$$

while the L-energy addition rule is given by

$$U^{(3)}_{q1,2} = U^{(3)}_{q1} + U^{(3)}_{q2}. \tag{12}$$

Notice that the addition rule of P-energy is different from that of L-energy. They have different physical meanings. P-energy is regarded as the sum of molecular kinetic energy, potential energy of molecules (long-range and short-range interactions), and even the radiation energy in the two-phase system [33], so it is nonadditive in mathematics. On the other hand, L-energy is only the sum of molecular kinetic energy and their potential energy with short-range interactions [33], so it is additive. These two internal energies are both realistic in physics and are related to each other by the Tsallis factor.

If an isolated two-phase nonextensive system arrives at Tsallis equilibrium, or the q-equilibrium, where Tsallis entropy has the maximum value, then we the q-entropy in (1) satisfies

$$\delta S_{q1,2} = c_{q2} \delta S_{q1} + c_{q1} \delta S_{q2} = 0, \tag{13}$$

and the P-energy satisfies the variation (see Appendix A),

$$\delta U^{(2)}_{q1,2} = c_{q2} \delta U^{(2)}_{q1} + c_{q1} \delta U^{(2)}_{q2} = 0, \tag{14}$$

but the L-energy satisfies the variation,

$$\delta U^{(3)}_{q1,2} = \delta U^{(3)}_{q1} + \delta U^{(3)}_{q2} = 0. \tag{15}$$

Now we may assume the addition rule of volume. If we assume that the variation in volume of the isolated two-phase nonextensive system satisfies

$$\delta V_{1,2} = c_{q2} \delta V_1 + c_{q1} \delta V_2, \tag{16}$$

and the total volume is unchanged, then based on the first law of the thermodynamics (7) and the P-energy variation (14), the q-entropy variation (13) can produce that

$$\delta S_{q1,2} = c_{q2}\delta U_{q1}^{(2)}\left(\frac{1}{T_{q1}} - \frac{1}{T_{q2}}\right) + c_{q2}\delta V_1\left[\frac{P_{q1}}{T_{q1}} - \frac{P_{q2}}{T_{q2}}\right]. \tag{17}$$

On the other hand, based on the first law of thermodynamics (8) and L-energy variation (15), the q-entropy variation (13) can also produce that

$$\delta S_{q1,2} = \delta U_{q1}^{(3)}\left[\frac{c_{q2}}{T_1} - \frac{c_{q1}}{T_2}\right] + c_{q2}\delta V_1\left[\frac{P_1}{T_1} - \frac{P_2}{T_2}\right]. \tag{18}$$

If the system is in the q-equilibrium, we have that $\delta S_{q1,2}=0$. By using

$$T_q = c_q T, \qquad P_q = c_q P, \tag{19}$$

we find that the above two equations (17) and (18) both lead to the same q-equilibrium condition, namely,

$$T_{q1} = T_{q2} \quad \text{and} \quad P_{q1} = P_{q2}. \tag{20}$$

But these conditions are obtained more directly by using (17) than by using (18), and the physical meaning in (17) is clearer than in (18). Therefore, hereinafter we can only use the P-set of formalism based on Eq.(7) to discuss the nonextensive thermodynamics.

The condition (20) shows that when a system is at the q-equilibrium, the physical temperature and the physical pressure are both homogeneous in the total system. If we use the condition (20) as a start, one can easily prove the assumption (18) in turn by the aid of equations (13), (14) and (7) (see Appendix B), showing the consistency of the thermodynamic method proposed in this paper.

As a short summary of this section, let us give some comments on the energy addition rule (6). Notice that there is a cross term of microstate energy level in this rule, which is consistent with the probability independent assumption [33], namely,

$$p_{ij} = p_i p_j. \tag{21}$$

The cross term shows the nonadditive property of (6), which directly reflects the nonextensivity of the systems considered, i.e. ordinarily called nonextensive systems. The existence of macroscopic energy (4) in the cross term shows the non-locality of complex nonextensive systems. It constructs a link between the microscopic energy and the total energy of the systems in visual way.

## 3. The stability criterion of a nonextensive system

Although a nonextensive system is indivisible in practice, we can imaginarily divide it into two subsystems marked also by "1" and "2" (the phases may be regarded as a special case of subsystems). Assuming that the nonextensive system is an isolated one and has been at a $q$-equilibrium, the sufficient and necessary condition for stability is as follows,

$$\Delta S_{q1,2} = \delta S_{q1,2} + \frac{1}{2}\delta^2 S_{q1,2} < 0. \tag{22}$$

At the $q$-equilibrium state, we have

$$\delta S_{q1,2} = c_{q2}\delta S_{q1} + c_{q1}\delta S_{q2} = 0. \tag{23}$$

Moreover, there is

$$\delta^2 S_{q1,2} = c_{q2}\delta^2 S_{q1} + c_{q1}\delta^2 S_{q2} + \frac{2(1-q)}{k}\delta S_{q1}\delta S_{q2}. \quad (24)$$

If we regard the subsystem 1 as the observed system, and 2 as the huge heat bath in canonical ensemble, the last two terms in above equation can be ignored. Also considering the positive definiteness of Tsallis factor, the stability criterion of the isolated nonextensive system becomes

$$\delta^2 S_{q1} < 0. \quad (25)$$

The importance of this criterion is that the stability of a given isolated system is determined by the stability of any part of this system. That is, once any part violates the stability criterion the whole system can not be stable.

According to (7), we have

$$\delta S_{q1} = \frac{1}{T_{q1}}(\delta U_{q1}^{(2)} + P_{q1}\delta V_1). \quad (26)$$

Regarding $\delta U_{q1}^{(2)}$ and $\delta V_1$ as the independent variables, that is,

$$\delta^2 U_{q1}^{(2)} = \delta^2 V_1 = 0, \quad (27)$$

we obtain that

$$\delta^2 S_{q1} = \delta(\delta S_{q1}) = \delta(\frac{1}{T_{q1}})\delta U_{q1}^{(2)} + \delta(\frac{P_{q1}}{T_{q1}})\delta V_1 = \frac{1}{T_{q1}}(-\delta T_{q1}\delta S_{q1} + \delta P_{q1}\delta V_1). \quad (28)$$

Furthermore, we can deduce that

$$-\delta T_{q1}\delta S_{q1} + \delta P_{q1}\delta V_1 = -\delta T_{q1}[(\frac{\partial S_{q1}}{\partial T_{q1}})_{V_1}\delta T_{q1} + (\frac{\partial S_{q1}}{\partial V_1})_{T_{q1}}\delta V_1]$$
$$+[(\frac{\partial P_{q1}}{\partial T_{q1}})_{V_1}\delta T_{q1} + (\frac{\partial P_{q1}}{\partial V_1})_{T_{q1}}\delta V_1]\delta V_1. \quad (29)$$

Considering the Maxwell relation, i.e.,

$$(\frac{\partial S_{q1}}{\partial V_1})_{T_{q1}} = (\frac{\partial P_{q1}}{\partial T_{q1}})_{V_1}, \quad (30)$$

we have that

$$-\delta T_{q1}\delta S_{q1} + \delta P_{q1}\delta V_1 = -(\frac{\partial S_{q1}}{\partial T_{q1}})_{V_1}(\delta T_{q1})^2 + (\frac{\partial P_{q1}}{\partial V_1})_{T_{q1}}(\delta V_1)^2. \quad (31)$$

Therefore, the stability criterion can be written as

$$\delta^2 S_{q1} = -\frac{C_{V1}}{T_{q1}^2}(\delta T_{q1})^2 + \frac{1}{T_{q1}}(\frac{\partial P_{q1}}{\partial V_1})_{T_{q1}}(\delta V_1)^2 < 0, \quad (32)$$

where the heat capacity with fixed volume is defined by

$$C_{V1} \equiv (\frac{\partial U_{q1}^{(2)}}{\partial T_{q1}})_{V_1} = T_{q1}(\frac{\partial S_{q1}}{\partial T_{q1}})_{V_1}. \quad (33)$$

Then the stability criterion becomes

$$C_{V1} > 0, \qquad (\frac{\partial P_{q1}}{\partial V_1})_{T_{q1}} < 0. \qquad (34)$$

As a simple application, let us consider a self-gravitating system consisting of gaseous molecules. For a small subsystem of this system, there is

$$\frac{1}{T_{q1}}(\frac{\partial P_{q1}}{\partial V_1})_{T_{q1}} \approx \frac{1}{T_1}(\frac{\partial P_1}{\partial V_1})_{T_1} < 0, \qquad (35)$$

where $P_1$ and $T_1$ represent the Lagrange pressure and temperature in the subsystem. That is to say, the subsystem behaves like a normal gas, whose isothermal compressibility is always positive. In this case, the stability of the self-gravitating system is determined by whether the heat capacity of any part satisfies

$$C_{V1} > 0. \qquad (36)$$

This is actually the necessary condition for stability of a self-gravitating system. It indicates that as long as the heat capacity of any part is negative, the self-gravitating system is always unstable.

The heat capacity in (36) is defined as the derivative of the physical internal energy to the physical temperature. Now that the physical temperature is the state variable defined directly through the zeroth law of thermodynamics, the heat capacity defined on basis of it should be the vital quantity for determining the thermodynamic evolution and stability of a given nonextensive system.

In a gaseous self-gravitating system, a new concept called the gravitational temperature (equivalent to the physical temperature in the ensemble theory) is introduced by using the molecular dynamic approach [34]. The gravitational heat capacity, defined as the derivative of the total energy to the gravitational temperature, can also be used to determine the stability of such systems. So in the classically gravitational thermodynamics [35], the specific heat, which is based on the classical temperature (it is identical to the Lagrange temperature in experiments), is irrelevant to the stability of the self-gravitating systems.

Silva R and Alcaniz J S deduced an expression for explaining the negative specific heat, in gaseous self-gravitating system [36]. However, the specific heat here is still based on the Lagrange temperature, and therefore its sign is not crucial to determine the stability of the system. The same case appears in the work of reference [37].

**4. The fluctuation theory in nonextensive thermodynamics**

Considering a nonextensive system is in contact with a heat reservoir, and the composite system is isolated. Assume when the microstate number is at the maximum, the energy and volume of the observed system are $\bar{E}$ and $\bar{V}$, respectively. Then the total entropy is written as

$$\bar{S}_{q(0)} = k \ln_q \bar{\Omega}_{(0)}, \qquad (37)$$

where $\Omega$ is the microstate number of system. When there are fluctuations $\Delta E$ and $\Delta V$ about the energy and volume of the system considered, the total entropy of the composite system becomes

$$S_{q(0)} = k \ln_q \Omega_{(0)}. \qquad (38)$$

According to the equiprobability principle, the probability $W$ of the considered system with the fluctuations $\Delta E$ and $\Delta V$ is

$$W \propto \frac{\Omega_{(0)}}{\overline{\Omega}_{(0)}} = \exp_q[\frac{\Delta S_{q(0)}}{k\overline{\Omega}_{(0)}^{1-q}}], \tag{39}$$

where

$$\Delta S_{q(0)} \equiv S_{q(0)} - \overline{S}_{q(0)}, \qquad \exp_q x \equiv [1+(1-q)x]^{1/(1-q)}. \tag{40}$$

Similar to (23), the deviation of the total Tsallis entropy is

$$\Delta S_{q(0)} = c_{qr}\Delta S_q + c_q \Delta S_{qr}, \tag{41}$$

where $\Delta S_q$ is the entropy variations of the system and $\Delta S_{qr}$ is that of the heat reservoir. According to (7), there is

$$\Delta S_{qr} = \frac{\Delta E_r + P_q \Delta V_r}{T_q}. \tag{42}$$

The quantity $E$ is recognized as the nonadditive energy in (3). That is because this energy is the sum of all the energy forms inside the system, including the possible long-range potential energy. Here, for convenience we employ the symbol '$E$' instead of that one in (3). $T_q$ and $P_q$ are respectively the physical temperature and the physical pressure of the heat reservoir, equivalent to those of the observed system at the $q$-equilibrium state. In view of the addition rule of energy and volume, namely,

$$\Delta E_{(0)} = c_{qr}\Delta E + c_q \Delta E_r = 0, \tag{43}$$

$$\Delta V_{(0)} = c_{qr}\Delta V + c_q \Delta V_r = 0, \tag{44}$$

we can get that

$$\Delta S_{q(0)} = c_{qr}\Delta S_q - c_{qr}\frac{\Delta E + P_q \Delta V}{T_q}. \tag{45}$$

Notice that

$$c_q = \Omega^{1-q}, \qquad \overline{\Omega}_{(0)}^{1-q} = c_q c_{qr}, \tag{46}$$

so we have that

$$W \propto \exp_q\left(-\frac{\Delta E - T_q \Delta S_q + P_q \Delta V}{kT_q c_q}\right). \tag{47}$$

Now we make a Taylor expansion for the system energy about its average value, and only consider the first order and second order terms, namely,

$$E = \overline{E} + (\frac{\partial E}{\partial S_q})_V \Delta S_q + (\frac{\partial E}{\partial V})_{S_q} \Delta V + \frac{1}{2}\frac{\partial^2 E}{\partial S_q^2}(\Delta S_q)^2$$
$$+ \frac{\partial^2 E}{\partial S_q \partial V}\Delta S_q \Delta V + \frac{1}{2}\frac{\partial^2 E}{\partial V^2}(\Delta V)^2. \tag{48}$$

Considering that

$$\Delta E = E - \bar{E}, \quad (\frac{\partial E}{\partial S_q})_V = T_q, \quad (\frac{\partial E}{\partial V})_{S_q} = -P_q, \tag{49}$$

we find that

$$\Delta E - T_q \Delta S_q + P_q \Delta V = \frac{1}{2} \frac{\partial^2 E}{\partial S_q^2} (\Delta S_q)^2 + \frac{\partial^2 E}{\partial S_q \partial V} \Delta S_q \Delta V + \frac{1}{2} \frac{\partial^2 E}{\partial V^2} (\Delta V)^2$$

$$= \frac{1}{2} (\Delta S_q \Delta T_q - \Delta P_q \Delta V). \tag{50}$$

Similar to (31), we have

$$\Delta S_q \Delta T_q - \Delta P_q \Delta V = \frac{C_V}{T_q} (\Delta T_q)^2 - (\frac{\partial P_q}{\partial V})_{T_q} (\Delta V)^2. \tag{51}$$

So equation (47) is changed into

$$W \propto \exp_q \left( -\frac{C_V}{2kT_q^2 c_q} (\Delta T_q)^2 + \frac{1}{2kT_q c_q} (\frac{\partial P_q}{\partial V})_{T_q} (\Delta V)^2 \right). \tag{52}$$

With this distribution function (52) of the fluctuations, if the average value is defined as

$$\overline{\Delta A} = \frac{\int W \Delta A \, d\Delta T_q \, d\Delta V}{\int W \, d\Delta T_q \, d\Delta V}, \tag{53}$$

where $\Delta A$ represents the deviation of any observable quantity $A$ from its average, there must be

$$\overline{\Delta T_q} = \overline{\Delta V} = 0. \tag{54}$$

Moreover, now that $\Delta T_q$ and $\Delta V$ are statistically independent, we can set $\Delta V = 0$ and can obtain (see the Appendix C) that

$$\overline{(\Delta T_q)^2} = \frac{\int \exp_q \left( -\frac{C_V}{2kT_q^2 c_q} (\Delta T_q)^2 \right) (\Delta T_q)^2 \, d\Delta T_q}{\int \exp_q \left( -\frac{C_V}{2kT_q^2 c_q} (\Delta T_q)^2 \right) d\Delta T_q} = \frac{2}{5-3q} \frac{kT_q^2 c_q}{C_V}. \tag{55}$$

Likewise, set $\Delta T_q = 0$, we obtain

$$\overline{(\Delta V)^2} = \frac{\int \exp_q \left( \frac{1}{2kT_q c_q} (\frac{\partial P_q}{\partial V})_{T_q} (\Delta V)^2 \right) (\Delta V)^2 \, d\Delta V}{\int \exp_q \left( \frac{1}{2kT_q c_q} (\frac{\partial P_q}{\partial V})_{T_q} (\Delta V)^2 \right) d\Delta V} = \frac{-2}{5-3q} kT_q c_q (\frac{\partial V}{\partial P_q})_{T_q}. \tag{56}$$

The ensemble theory demands that the fluctuations in (55) and (56) must be more than zero. This requires the parameter to be q<5/3, a restriction to value of the nonextensive parameter *q*. Regarding the energy *E* as the function of independent variables $T_q$ and $V$, we can get

$$\Delta E = (\frac{\partial E}{\partial T_q})_V \Delta T_q + (\frac{\partial E}{\partial V})_{T_q} \Delta V. \tag{57}$$

Therefore, we have that

$$\overline{(\Delta E)^2} = C_V^2 \overline{(\Delta T_q)^2} + \left(\frac{\partial E}{\partial V}\right)_{T_q}^2 \overline{(\Delta V)^2}$$

$$= \frac{2}{5-3q}\left[kT_q^2 c_q C_V - kT_q c_q (\frac{\partial V}{\partial P_q})_{T_q} \left(\frac{\partial E}{\partial V}\right)_{T_q}^2\right]. \tag{58}$$

This result is different from that one deduced in the energy representation of (4) [38-39]. However, when $q\to 1$, the energy fluctuation (58) also recovers the classical result, namely,

$$\overline{(\Delta E)^2} = kT^2 C_V - kT(\frac{\partial V}{\partial P})_T \left(\frac{\partial E}{\partial V}\right)_T^2. \tag{59}$$

**5. Summary and conclusion**

In this work, we have studied the balance conditions of a two-phase nonextensive system. By assuming that volume is nonadditive, we find that it is the physical pressure that is responsible for the mechanical balance between the two phases. This nonadditive assumption is that the volume of one phase existing individually is different from that of the phase mixing with the other phase. The basic reason is because a nonexensive system is indivisible in practice. Here, the assumption of temperature duality plays a fundamental role. The balance conditions can be obtained in both the two sets of thermodynamic formalisms. In the P-set, the inference about the balance conditions is more direct and convenient, therefore the discussion of thermodynamic issues is carried out ordinarily in the P-set formalism.

For an isolated system, the sufficient and necessary condition of stability is that the virtual change in the total entropy is less than zero, see (22). Then the inequality (25) is deduced, which shows that *the stability of the whole system is essentially determined by the stability of any given part of it*. Through the mathematical transformation, the stability criterion (34) can be expressed by the heat capacity and the isothermal compressibility. This criterion might be important in studying self-gravitating gaseous system. In such a system, the isothermal compressibility is always positive. So the stability of the whole self-gravitating system is determined by whether the heat capacity of any part is positive. That is to say, only when the heat capacity of the system's core is positive, can the whole system be stable. Otherwise, the self-gravitating system is unstable.

In particular, using the method of molecular dynamics, one can define an concept named the gravitational temperature [34], which is equivalent to the physical temperature defined in ensemble theory. The gravitational heat capacity defined by the gravitational temperature plays a vital role in the stability and evolution of a self-gravitating system. Other definitions of specific heat in Refs.[35-37] are all based on the Lagrange temperature, and are irrelevant to the stability of such systems.

The quasi-thermodynamic theory of fluctuation is generalized to the nonextensive systems. We obtain the generalized form (39) of the fluctuation in the entropy. Based on this formula, we can easily calculate the probability of the fluctuations in the physical temperature and volume. The inferred energy fluctuation (58) is different from that one obtained in the representation of energy (4). The generalized fluctuation theory given here might be useful to study the critical fluctuation phenomena where the nonextensivity can not be ignored.

The fundamental treatment in this paper is the dual explanation of temperature [31], based on which a thermodynamic formalism consisting of two parallel Legendre structures is proposed [33]. Our treatment, along with the nonadditive composition rule of basic quantities such as entropy,

P-energy and volume, is consistent with the zeroth law of thermodynamics, to a certain extent which resolves the fundamental controversy about the definition of temperature within the nonextensive thermodynamics.

There was another method called formal logarithm [40] to resolve the same controversy. In the light of this method, the nonadditive composition rule existing generally in nonextensive statistics is mapped to an additive composition rule, where the basic quantities are redefined by the formal logarithm. This prescription made another attempt to construct the nonextensive thermodynamic formalism which can be applicable in the complex systems. With this prescription, the thermodynamic problem in black hole entropy was reexamined [41], where the new defined additive entropy is proportional to the energy of a black hole, and the derived temperature is a constant for all black holes.

**Acknowledgements**

This work is supported by the National Natural Science Foundation of China under Grants No. 11405092 and No. 11775156.

**Appendix A**

In order to prove (16), we need (13), namely,

$$U_{q1,2}^{(2)} = c_{q2}U_{q1}^{(2)} + c_{q1}U_{q2}^{(2)} - (1-q)\beta' U_{q1}^{(2)} U_{q2}^{(2)}, \tag{A.1}$$

In realm of the energy representation (3), there is the following identity [33],

$$c_q = [Z_q^{(2)}]^{1-q} + (1-q)\beta' U_q^{(2)}. \tag{A.2}$$

The quantity $Z_q^{(2)}$ is the partition function defined in representation (3), which is analytical in mathematics. It is the function of the generalized Lagrange multiplier $\beta'$. At q-equilibrium state, the variation in the generalized Lagrange multiplier is zero. This means

$$\delta c_q = (1-q)\beta' \delta U_q^{(2)}. \tag{A.3}$$

So we have

$$\delta[c_{q2} U_{q1}^{(2)}] = c_{q2} \delta U_{q1}^{(2)} + (1-q)\beta' \delta U_{q2}^{(2)} U_{q1}^{(2)}, \tag{A.4}$$

$$\delta[c_{q1} U_{q2}^{(2)}] = c_{q1} \delta U_{q2}^{(2)} + (1-q)\beta' \delta U_{q1}^{(2)} U_{q2}^{(2)}. \tag{A.5}$$

Therefore, for an isolated system, when it arrives at equilibrium, there should be that

$$\begin{aligned}\delta U_{q1,2}^{(2)} &= c_{q2}\delta U_{q1}^{(2)} + (1-q)\beta'\delta U_{q2}^{(2)} U_{q1}^{(2)} + c_{q1}\delta U_{q2}^{(2)} + (1-q)\beta'\delta U_{q1}^{(2)} U_{q2}^{(2)} \\ &\quad -(1-q)\beta'\delta U_{q2}^{(2)} U_{q1}^{(2)} - (1-q)\beta'\delta U_{q1}^{(2)} U_{q2}^{(2)} \\ &= c_{q2}\delta U_{q1}^{(2)} + c_{q1}\delta U_{q2}^{(2)} \\ &= 0\end{aligned} \tag{A.6}$$

The equation (16) is then proved.

**Appendix B**

In order to prove (18), let us rewrite down the equations (15), (16) and (7), namely,

$$\delta S_{q1,2} = c_{q2}\delta S_{q1} + c_{q1}\delta S_{q2}, \tag{B.1}$$

$$\delta U_{q1,2}^{(2)} = c_{q2}\delta U_{q1}^{(2)} + c_{q1}\delta U_{q2}^{(2)}, \tag{B.2}$$

$$\delta U_q^{(2)} = T_q \delta S_q - P_q \delta V. \tag{B.3}$$

Substituting (B.3) into (B.2), we have,

$$\begin{aligned} & T_{q1,2}\delta S_{q1,2} - P_{q1,2}\delta V_{1,2} \\ & = c_{q2}[T_{q1}\delta S_{q1} - P_{q1}\delta V_1] + c_{q1}[T_{q2}\delta S_{q2} - P_{q2}\delta V_2] \end{aligned}. \tag{B.4}$$

Now that the system considered is at equilibrium, considering the equilibrium condition (21), there should be that

$$T_{q1,2} = T_{q1} = T_{q2} \equiv T_q, \tag{B.5}$$

$$P_{q1,2} = P_{q1} = P_{q2} \equiv P_q. \tag{B.6}$$

Furthermore, substituting (B.1) into (B.4), one gets,

$$-P_q \delta V_{1,2} = -P_q c_{q2} \delta V_1 - P_q c_{q1} \delta V_2. \tag{B.7}$$

Removing the minus physical pressure on both sides of the above equation, one immediately obtains (18).

On the other hand, using the equation (17), we have that

$$\begin{aligned} & T_{1,2}\delta S_{q1,2} - P_{1,2}\delta V_{1,2} \\ & = T_1 \delta S_{q1} - P_1 \delta V_1 + T_2 \delta S_{q2} - P_2 \delta V_2 \end{aligned}. \tag{B.8}$$

By the aid of the relation,

$$c_{q1,2} = c_{q1}c_{q2}, \tag{B.9}$$

and using $T_q = c_q T$ and $P_q = c_q P$, we also can derive equation (B.4), and then obtain (18). The equation (18) is then proved. The above proof shows that this addition rule (18) of the volume is irrelevant to the choice of the definition of the energy.

**Appendix C**

In order to prove (57), let us introduce that

$$I_1 \equiv \int \exp_q[-\frac{C_V}{2kT_q^2 c_q}(\Delta T_q)^2](\Delta T_q)^2 d\Delta T_q, \tag{C.1}$$

$$I_2 \equiv \int \exp_q[-\frac{C_V}{2kT_q^2 c_q}(\Delta T_q)^2] d\Delta T_q. \tag{C.2}$$

Then we have that

$$\overline{(\Delta T_q)^2} = \frac{I_1}{I_2}. \tag{C.3}$$

Now considering that

$$I_1 = 2\int_0^a [1-(1-q)\frac{C_V}{2kT_q^2 c_q}(\Delta T_q)^2]^{\frac{1}{1-q}} (\Delta T_q)^2 d\Delta T_q$$

$$= \frac{kT_q^2 c_q}{-C_V} \frac{2}{2-q} \int_0^a d[1-(1-q)\frac{C_V}{2kT_q^2 c_q}(\Delta T_q)^2]^{\frac{1}{1-q}+1} \Delta T_q$$

$$= \frac{kT_q^2 c_q}{C_V} \frac{2}{2-q} \int_0^a [1-(1-q)\frac{C_V}{2kT_q^2 c_q}(\Delta T_q)^2]^{\frac{1}{1-q}+1} d\Delta T_q, \quad \text{(C.4)}$$

where,

$$a = \begin{cases} \sqrt{\dfrac{2kT_q^2 c_q}{(1-q)C_V}}, & 1-q > 0 \\ \infty, & 1-q < 0 \end{cases}, \quad \text{(C.5)}$$

one can find that

$$I_1 = \frac{kT_q^2 c_q}{C_V} \frac{1}{2-q} [2\int_0^a [1-(1-q)\frac{C_V}{2kT_q^2 c_q}(\Delta T_q)^2]^{\frac{1}{1-q}} [1-(1-q)\frac{C_V}{2kT_q^2 c_q}(\Delta T_q)^2] d\Delta T_q]$$

$$= \frac{kT_q^2 c_q}{C_V} \frac{1}{2-q} [I_2 - (1-q)\frac{C_V}{2kT_q^2 c_q} I_1] = \frac{kT_q^2 c_q}{C_V} \frac{1}{2-q} I_2 - \frac{1-q}{2(2-q)} I_1. \quad \text{(C.6)}$$

From the above expression, it is easy to obtain the fluctuation,

$$\overline{(\Delta T_q)^2} = \frac{I_1}{I_2} = \frac{2}{5-3q} \frac{kT_q^2 c_q}{C_V}. \quad \text{(C.7)}$$

So equation (57) is verified. This approach is similar to that in Ref.[39].

**References**

[1] Tsallis, C. J. Stat. Phys. 52, 479 (1988).
[2] Lima, J.A.S., Silva Jr., R., Santos, J. Phys. Rev. E 61, 3260 (2000).
[3] Du, J.L. Phys. Lett. A 329, 262 (2004).
[4] Liu, L.Y., Du, J.L. Physica A 387, 4821 (2008)
[5] Yu, H.N., Du, J.L. Ann. Phys. 350, 302 (2014).
[6] Du, J.L. Europhys. Lett. 67, 893 (2004)
[7] Leubner, M.P. Astrophys. J. 632, L1 (2005)
[8] Du, J.L. New Astron. 12, 60 (2006)
[9] Du, J.L. Astrophys. Space Sci. 312, 47 (2007) and the references therein
[10] Liu, B., Goree, J. Phys. Rev. Lett. 100, 055003 (2008)
[11] Liu, B., Goree, J., Feng, Y. Phys. Rev. E 78, 046403 (2008)
[12] Oikonomou, T., Provata, A., Tirnakli, U. Physica A 387, 2653 (2008)
[13] Rolinski, O.J., Martin, A., Birch, D.J.S. Ann. N. Y. Acad. Sci. 1130, 314 (2008)
[14] Eftaxias, K., Minadakis, G., Potirakis, S.M., Balasis, G. Physica A 392, 497 (2013)
[15] Anteneodo, C., Tsallis, C., Martinez, A.S. Europhys. Lett. 59, 635 (2002)
[16] Yamano, T. Eur. Phys. J. B 38, 665 (2004)
[17] Abe, S., Martınez, S., Pennini, F., Plastino, A. Phys. Lett. A 281, 126 (2001)
[18] Toral R. Physica A, 2003, 317(1): 209-212.
[19] Wang Q A, Nivanen L, Le Mehaute A, et al. EPL (Europhysics Letters), 2004, 65(5): 606.

[20] Abe S. Physica A 2006, 368(2): 430-434.
[21] Du, J.L. Bull. Astr. Soc. India 35, 691 (2007)
[22] Guo, L.N., Du, J.L. Physica A 388, 4936 (2009)
[23] Guo, L.N., Du, J.L. Physica A 390, 183 (2011)
[24] Ou C., Chen J. Physica A 370, 525 (2006).
[25]Abe, S. Physica A 269(2), 403–409 (1999)
[26] Abe, S. Phys. Rev. E 63, 061105 (2001)
[27] Martınez, S., Pennini, F., Plastino, A. Physica A 295, 416 (2001)
[28] Johal, R.S. Phys. Lett. A 318, 48 (2003)
[29] Johal, R.S. Phys. Lett. A 332, 345 (2004)
[30] Scarfone, A.M. Phys. Lett. A 374, 2701 (2010)
[31] Zheng, Y.H. Physica A 392, 2487 (2013)
[32] Zheng, Y.H., Du, J.L. Physica A 427, 113 (2015)
[33] Zheng Y, Du J., J. Continuum Mech. Thermodyn. (2016) 28: 1791.
[34] Zheng Y, Du J. Physica A 420 (2015) 41-48.
[35] Lynden-Bell, Donald. Physica A 263.1-4 (1999): 293-304.
[36] Silva R, Alcaniz J S. Physics Letters A, 2003, 313(5): 393-396.
[37] Taruya A, Sakagami M. Physica A, 2002, 307(1): 185-206.
[38] Liu L, Du J. Physica A 2008, 387(22): 5417-5421.
[39] Guo R, Du J. Physica A 2012, 391(9): 2853-2859.
[40] Ván P, Barnaföldi G G, Biró T S, et al. Journal of Physics: Conference Series. IOP Publishing, 2012, 394(1): 012002.
[41] Czinner, Viktor G. International Journal of Modern Physics D 24.09 (2015): 1542015.